# A Novel Unsupervised Graph Wavelet Autoencoder for Mechanical System Fault Detection

Tianfu Li[a, *], Chuang Sun[b, *], Ruqiang Yan[b], Xuefeng Chen[b]

a. Faculty of Mechanical and Electrical Engineering, Kunming University of Science and Technology, Kunming, 650550, China

b. School of Mechanical Engineering, Xi'an Jiaotong University, Xi'an, Shaanxi 710049, China.

E-mail address: tianfu.li@ieee.org, ch.sun@xjtu.edu.cn, yanruqiang@xjtu.edu.cn, chenxf@xjtu.edu.cn, *Corresponding author.

**Abstract:**

Reliable fault detection is an essential requirement for safe and efficient operation of mechanical systems in various industrial applications. Despite the abundance of existing approaches and the maturity of fault detection research field, the interdependencies between condition monitoring data have often been overlooked. Recently, graph autoencoder (GAE) have been proposed as a solution for learning the interdependencies among data and has gained widespread use in fault detection. However, both GAE and graph variational autoencoder (GVAE) have fixed receptive fields, limiting their ability to extract multiscale features and model performance. To overcome these limitations, we propose two graph neural network models: the graph wavelet autoencoder (GWAE), and the graph wavelet variational autoencoder (GWVAE). GWAE consists mainly of the spectral graph wavelet convolutional (SGWConv) encoder and a feature decoder, while GWVAE is the variational form of GWAE. To achieve unsupervised mechanical system fault detection, we transform the collected system signals into PathGraph by considering the neighboring relationships of each data sample. Fault detection is then achieved by evaluating the reconstruction errors of normal and abnormal samples. Three case studies are carried out to verify the effectiveness of the proposed methods, which obtains performance improvements by around 3%~4% compared to the comparison methods.

# 1  Introduction

Early fault detection (FD) is of great significance for prognostics and health management (PHM) of mechanical equipment and serves as the basis for equipment condition monitoring [1-3]. FD is the process of identifying anomalies or deviations from normal operating conditions in a system or a component. This process typically requires monitoring the signals generated by condition monitoring sensors and identifying data samples that deviate from the normal operating conditions [4-6]. Timely and accurate FD of equipment can prevent damage, effectively reducing equipment downtime and minimizing losses caused by failures [7]. The recent collection of vast amounts of data on the condition and operation of the systems presents both new opportunities and challenges for the early FD of mechanical systems [8-10].

Currently, there exists a wide range of FD methods, with deep learning (DL)-based approaches being extensively employed in the field [11, 12]. These methods have the capability to automatically extract fault information from different types of data, including the high-frequency vibration and sound signals [13]. Among these DL-based FD methods, unsupervised anomaly detection (UAD) approaches have gained prominence as they can learn features from complex data in an unsupervised manner, bypassing the need for labeled data [14]. At present, there are two commonly applied UAD approaches: autoencoder (AE)-based methods [15] and one-class classification methods (OCCs) [16, 17]. For the AE-based methods, when trained on healthy data, their reconstruction errors can be used to distinguish between healthy and faulty system conditions. Several types of AEs have been proposed, including convolutional autoencoders (CAEs) [18], and convolutional variational autoencoders (CVAE) [19]. While OCC methods provide a unified way to making decisions on system anomalies of the by establishing a decision boundary of the health state, any data outside this boundary is considered abnormal. Notable OCC methods include the Support Vector Data Description (SVDD) and its extension, DeepSVDD, both of which aim to find a hypersphere in the feature space that encloses the normal data points, considering data outside the hypersphere as anomalies [20].

SVDD uses a kernel function to map the input data into a high-dimensional feature space, while DeepSVDD employs a deep neural network to learn a compact representation of the input data. DeepSVDD learns the center and radius of the hypersphere in the learned feature space.

Despite the notable achievements of unsupervised FD methods, there is a tendency to overlook the significance of capturing dynamically evolving temporal relationships at various time scales during the feature extraction process. Recently, graph neural networks (GNNs) have evolved as a solution for modelling graph-structured data and capturing the spatial-temporal relationships in multi-variate time series data [21], utilizing techniques such as graph convolutional networks (GCNs) [22] and graph attention networks (GATs) [23]. Furthermore, graph autoencoders (GAE) and graph variational encoders (GVAE) have been applied to FD problems in mechanical systems [24]. The concepts are similar to those of standard AEs: utilizing reconstruction errors for performing FD [21]. More recently, GCNs with attention modules [25] and graph transformers [26] were proposed as encoders and have also demonstrated promising results for FD in mechanical systems. However, existing FD methods based on GAE and GVAE are limited by their fixed receptive field. This limitation restricts their ability to extract long-ranging temporal multiscale relationships, which in turn affects their capability to learn more informative feature representations.

In this paper, we introduce a novel unsupervised framework for FD in mechanical systems based on graph AEs. Our approach includes two variants: graph wavelet autoencoder (GWAE) and graph wavelet variational autoencoder (GWVAE). GWAE consists of a spectral graph wavelet convolutional (SGWConv) encoder and a two-layer perceptron decoder, while GWVAE is the variational counterpart of GWAE. The SGWConv encoder is built upon the spectral graph wavelet transform, which enables effective multiscale feature extraction by decomposing the graph signals into scaling function coefficients and spectral graph wavelet coefficients. To evaluate the performance of our method, we conduct experiments using two condition monitoring datasets collected from fuel control systems and one acoustic monitoring dataset from a valve. The contributions of this paper can be summarized as follows:

1) We develop the SGWConv encoder based on the spectral graph wavelet transform, which realizes multiscale feature extraction with a low-pass filter and several band-pass filters.

2) We construct GWAE and GWVAE based on the SGWConv encoder and two-layer perceptron decoder, which obtain anomaly scores in an unsupervised manner.

3) By transforming the raw signals into the PathGraph representation, we consider the long-ranging temporal multi-scale relationships in the high-frequency data and realize FD using the proposed GWAE and GWVAE.

The remainder of this work is organized as follows: Section 2 reviews the preliminaries of GCNs and GAEs. Section 3 presents the detailed description of the proposed methods. In Section 4, we evaluate the performance of the proposed framework on two datasets of the fuel control system. Additionally, in Section 5, we utilize one acoustic monitoring dataset from a valve to further validate the proposed methods. Finally, the conclude our findings in Section 6.

## 2 Preliminary

### 2.1 Graph Convolutional Networks

Given a graph G=($N$, $\mathbf{X}$, $\mathbf{A}$), where $N$ is the number of nodes, $\mathbf{X} \in \mathrm{R}^{N \times d}$ is the feature matrix of the graph with $d$ denoting the feature dimension of a node, and $\mathbf{A} \in \mathrm{R}^{N \times N}$ is the adjacency matrix of the graph. Specifically, if $\mathbf{A}_{i,j}=1$, there is an edge between nodes $i$ and $j$, and if $\mathbf{A}_{i,j}=0$, the two nodes are not connected.

Currently, the most broadly used GNNs can be categorized into two main types: spectral GCNs and spatial GCNs [27]. The standard spectral graph convolution is derived from the graph Fourier transform, which is defined as follows:

$$\mathbf{X} *_{G} \mathbf{g}(\theta) = \mathbf{U} \mathbf{g}(\theta) \mathbf{U}^{\mathrm{T}} \mathbf{X} \qquad (1)$$

where $\mathbf{g}(\theta)$ is the learnable matrix, $\mathbf{U}$ represents the eigenvector of graph Laplacian matrix $\mathbf{L}$ defined as $\mathbf{L}=\mathbf{D}\text{-}\mathbf{A}$, and $\mathbf{D}$ is diagonal matrix with $\mathbf{D}_{i,i}=\sum_j \mathbf{A}_{i,j}$. Additionally, $\mathbf{U}^T\mathbf{X}$ denotes the graph Fourier transform here.

Spatial GCNs are defined directly in the vertex domain and typically involve three steps: 1) aggregating the node features of adjacent nodes, 2) updating the node features of the central node with the aggregated node features, and 3) obtaining the representations of the entire graph with the readout operator [28]. Consequently, the feature representations of a graph after undergoing these processes can be formalized as follows:

$$\mathbf{h}_{v_i}^k = \text{Update}(\text{Aggregate}(\mathbf{h}_{v_j}^{k-1}), \mathbf{h}_{v_i}^{k-1}) \tag{2}$$

$$\mathbf{h}_G = \text{Readout}(\mathbf{h}_v^k) \tag{3}$$

where $v_j \in N(v_i)$ represents the adjacent nodes of $v_i$ and $\mathbf{h}_{v_j}^{k-1}$ corresponds to the node features of $v_j$. $\mathbf{h}_{v_i}^k$ represents the updated node features of $v_i$ at the $k$-th iteration, and $\mathbf{h}_G$ denotes the graph representation. In general, Aggregate(·) and Readout(·) operators often utilize mean, max, or sum operations, while the Update(·) operator commonly employs concatenation or addition operations.

## 2.2 Graph Autoencoders

GAE [29] is developed as an analogy to traditional AE, enabling unsupervised learning of the graph structured data. However, unlike the symmetric structure of AE, GAE typically consists of two decoders: one for feature reconstruction and another for graph structure reconstruction.

The original GAE and GVAE [30] employed the inner product of the learned latent representation as the model decoder for reconstructing the adjacency matrix. The reconstructed adjacency matrix, denoted as $\hat{\mathbf{A}}$, can be represented as:

$$\hat{\mathbf{A}} = \text{Sigmoid}(\mathbf{Z} \cdot \mathbf{Z}^T) \tag{4}$$

where $\mathbf{Z}$ denotes the latent representation learned by the graph encoder and Sigmoid(·) denotes the activation function. The models are then trained based on the mean squared error (MSE) between the true and the reconstructed adjacency matrix.

However, since the inner product decoder is not learnable, an alternative approach is employed to address this limitation. The graph convolutional (GConv) layer or multilayer perceptron (MLP) are utilized to reconstruct the feature matrix [31]. Among them, GConv can use structural information during reconstruction, while MLP cannot. The reconstructed feature matrix can be expressed as

$$\hat{\mathbf{X}} = f(\mathbf{X}, \mathbf{A}) \text{ or } f(\mathbf{X}) \qquad (5)$$

where $\hat{\mathbf{X}}$ represents the reconstructed feature matrix, and $f(\cdot)$ denotes the feature decoder, which can take $f(\mathbf{X}, \mathbf{A})$ as input for the GConv layer, while the MLP uses $f(\mathbf{X})$ as input. The model can be updated using the objective function (e.g., mean absolute error or the mean squared error) between the two feature matrices. However, a study conducted in [32] discovered that using the graph convolutional layer as the encoder may not be suitable due to the Laplacian smoothing effect. Additionally, it is also feasible to simultaneously reconstruct the adjacency matrix $\mathbf{A}$ and feature matrix $\mathbf{X}$ [33], which can make the learned model more robust.

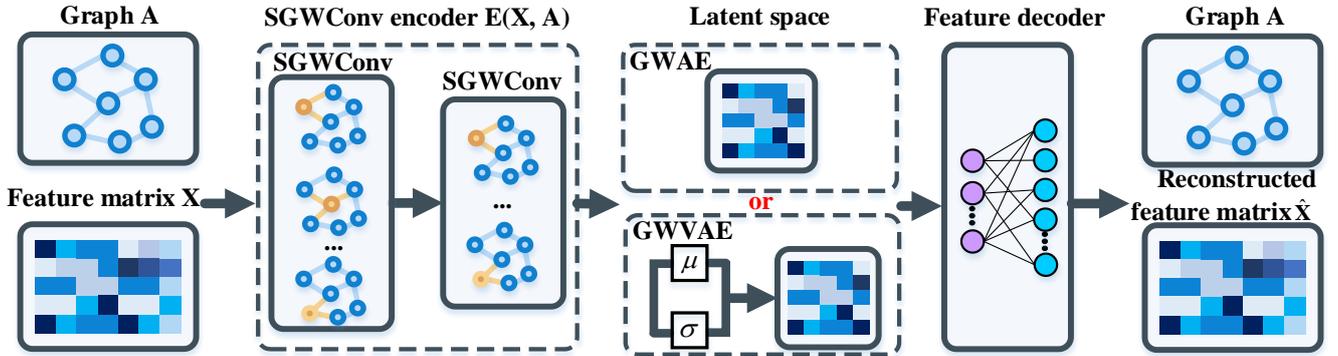

Fig. 1. The structure of the proposed GWAE and GWVAE.

## 3  Methodology

The framework of the proposed GWAE and GWVAE is illustrated in Fig. 1, which primarily comprises an SGWConv encoder and a two-layer perceptron decoder. In GWAE, the node feature matrix is directly reconstructed from the latent space, while GWVAE samples the learned data distribution to obtain the latent space. The expression for each component is explained in detail in the following.

## 3.1 Spectral Graph Wavelet Convolutional Encoder

The receptive field of MPNN is fixed, and its propagation matrices, which multiply graph signal, can be understood as performing low-pass filtering on the graph [34]. In order to achieve multiscale feature extraction, we introduce the spectral graph wavelet convolutional layer (SGWConv) based on the spectral graph wavelet transform. The SGWConv layer is defined as follows:

$$\mathbf{x} *_G \mathbf{g} = \mathbf{P}^T \mathbf{g}(\theta) \mathbf{P} \mathbf{x} \tag{6}$$

where $\mathbf{Px}$ represents the spectral graph wavelet transform (SGWT), $\mathbf{g}(\theta)$ is the learnable filter matrix in the wavelet domain, and $\mathbf{P}=[u(\mathbf{L}), v(s_1\mathbf{L}), v(s_2\mathbf{L}),\dots,v(s_J\mathbf{L})]$ represents the wavelet operator. The function $u(\cdot)$ serves as the scaling kernel function, which acts as a low-pass filter. Moreover, $v(\cdot)$ represents the wavelet kernel function, which forms scale-dependent band-pass filters. The set $S=\{s_1, s_2,\dots, s_J\}$ denotes the $J$ decomposition scales. The wavelet kernel function at different scales forms a set of band-pass filters, while the scaling kernel function acts as a low-pass filter.

In SGWT, various kernel functions can be chosen. In this case, we adopt the Mexican Hat wavelet and its corresponding scaling function as the kernel functions for the wavelet operator [35]. They are defined as

$$\begin{cases} u(\lambda) = \gamma e^{-Q\lambda/0.6\lambda_{max}} \\ v(s_j\lambda) = s_j\lambda e^{-s_j\lambda} \end{cases} \tag{7}$$

where $\lambda_{max}$ is the maximum Laplacian eigenvalue and $Q$ is a hyperparameter. $\gamma$ is defined as $v_{max}$ and is set to $e^{-1}$ in this case. Additionally, the detailed feature extraction process of SGWConv is illustrated in Fig. 2. In this process, the input graph signal is initially decomposed into scaling function coefficients and several graph wavelet kernel coefficients using the wavelet operator. Subsequently, the extracted multiscale coefficients are filtered with a graph filter, and the output signal is obtained by multiplying a reverse wavelet operator.

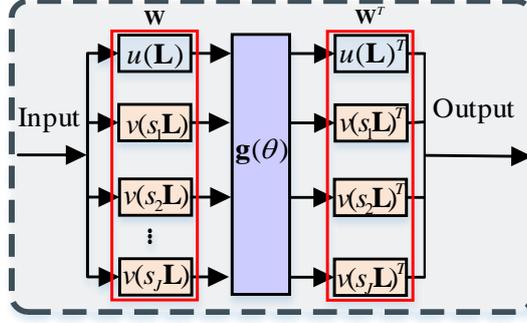

Fig. 2. The illustration of SGWConv.

The SGWConv encoder consists of two SGWConv layers, which are defined as follows:

$$\mathbf{Z}^{(1)} = \text{SGWConv}(\mathbf{X},\mathbf{A}) = f_{\text{ReLU}}(\mathbf{P}^T \mathbf{g}(\theta_1)\mathbf{P}\mathbf{X}) \quad (8)$$

$$\mathbf{Z}^{(2)} = \text{SGWConv}(\mathbf{X},\mathbf{A}) = f_{\text{ReLU}}(\mathbf{P}^T \mathbf{g}(\theta_2)\mathbf{P}\mathbf{Z}^{(1)}) \quad (9)$$

where $f_{\text{ReLU}}$ is the non-linear activation function, and $\mathbf{Z}$ denotes the learned hidden representations. Therefore, the process of using the encoder to encode the graph structure and node features into representations is denoted as $\mathbf{Z}=q(\mathbf{Z}|\mathbf{X},\mathbf{A})=\mathbf{Z}^{(2)}$.

For GWAVE, the encoder is an inference model, which can be mathematically defined as follows:

$$q(\mathbf{Z}|\mathbf{X},\mathbf{A}) = \prod_{i=1}^{N} q(\mathbf{z}_i|\mathbf{X},\mathbf{A}) \quad (10)$$

$$q(\mathbf{z}_i|\mathbf{X},\mathbf{A}) = \mathbb{N}(\mathbf{z}_i|\mathbf{\mu}_i,\text{diag}(\mathbf{\sigma}_i^2)) \quad (11)$$

where $\mathbf{\mu}_i=\text{SGWConv}_{\mathbf{\mu}}(\mathbf{X},\mathbf{A})$ is the matrix of mean vectors $\mathbf{z}_i$, and similarly, we can obtain $\log\mathbf{\sigma} = \text{SGWConv}_{\mathbf{\sigma}}(\mathbf{X},\mathbf{A})$.

## 3.2 Feature Decoder

In the context of a GAE, there are typically three options for reconstruction: reconstructing features $\mathbf{X}$, reconstructing structure $\mathbf{A}$, or reconstructing both $\mathbf{X}$ and $\mathbf{A}$. In this case, our reconstruction strategy focuses on reconstructing the feature matrix $\mathbf{X}$ of the graph data. By using a two-layer perceptron as the feature decoder, we aim to accurately capture the variations in sensor measurements caused by system anomalies. Therefore, the reconstructed node feature matrix can be represented as follows:

$$\hat{\mathbf{X}} = D(\mathbf{X}|\mathbf{Z}) = f_{\text{ReLU}}(\text{FC}(\mathbf{Z})) \quad (12)$$

where $\text{FC}(\cdot)$ represents a two-layer MLP, and $\hat{\mathbf{X}}$ denotes the reconstructed feature matrix.

Once the structure of the feature decoder is defined, the vanilla GWAE can be trained with a reconstruction loss, denoted as:

$$L_{RC} = \|\mathbf{X} - \hat{\mathbf{X}}\|_F^2 \tag{13}$$

where $\|\cdot\|$ is the Frobenius norm.

For the GWVAE, the loss function is defined as follows:

$$L_1 = L_{RC} + L_{KL} = \|\mathbf{X} - \hat{\mathbf{X}}\|_F^2 + \lambda \mathrm{KL}(q(\mathbf{Z}|\mathbf{X},\mathbf{A})\|p(\mathbf{Z})) \tag{14}$$

where $p(\mathbf{Z})$ is a Gaussian prior defined as $p(\mathbf{Z}) = \prod_i p(\mathbf{z}_i) = \prod_i \mathbb{N}(\mathbf{z}_i|0,\mathbf{I})$ with $\mathbf{I}$ being the identity matrix. $\mathrm{KL}(\cdot)$ represents the Kullback-Leibler divergence between $q(\cdot)$ and $p(\cdot)$, $\lambda$ is a tradeoff parameter that is set to 0.5 in this study, following the settings proposed in [30].

### 3.3 Anomaly Scores for Fault detection

Once the model is trained, it is able to reconstruct normal samples well, enabling it to detect faults. We define the anomaly score as follows:

$$\xi_i = \|\mathbf{x}_i - \hat{\mathbf{x}}\|_F^2 \tag{15}$$

To identify anomalous samples, those with high anomaly scores are considered as anomalies. However, the effectiveness of FD results often depends on selecting an appropriate threshold for identifying abnormal data. To mitigate the reliance on manual threshold selection in fault detection, this paper uses kernel density estimation (KDE) [36] to model the distribution of anomaly score from the training samples. A decision threshold is then determined at a specific significance level, ensuring a more automated and objective approach to threshold determination.

Given the anomaly scores of training samples, denoted as $\{\xi_i\}$, the probability density function $K(\xi)$ estimated using KDE is defined as follows:

$$K(\xi) = \frac{1}{MH} \sum_{i=1}^{M} F(\frac{\xi - \xi_i}{H}) \tag{16}$$

where $F(\cdot)$ is the kernel function of KDE with the radial basis function being adopted in this case. $M$

represents the number of training samples, and H is the bandwidth determined by $(\frac{M(d+2)}{4})^{-\frac{2}{d+4}}$[37], where $d$ denotes the sample dimension.

With the estimated $K(\xi)$, the cumulative distribution function CF($\xi$) can be calculated by CF($\xi$)=$\int_{\infty}^{\xi} K(\xi)d\xi$. Therefore, given the CF($\xi$) and a significance level $\delta \in [0,1]$, the decision threshold $\xi_\delta$ can be determined by satisfying the following condition:

$$\text{CF}(\xi_\delta) = 1 - \delta \qquad (17)$$

In this case, a sample with an anomaly score $\xi \geq \xi_\delta$ has a probability of at least $(1 - \delta)100\%$ of being an outlier. Since the significance level $\delta$ is a hyperparameter usually determined based on the costs of false positives and false negatives, and since we do not have costs defined in this research, we empirically set $\delta$ to 0.1 for fault detection.

### 3.4 Fault Detection with the Proposed Models

The loss functions of GWAE and GWVAE are presented in Eq. (13) and Eq. (14), respectively. The training process follows the same procedure as traditional AE and VAE. Once the model is trained, we determine the threshold for FD using Eq. 18. To summarize the entire FD procedure **Algorithm 1** is provided.

---

**Algorithm 1**: The FD procedure of GWAE and GWVAE.

**Inputs:** training set {$\mathbf{X}_{\text{train}}$}, validation set {$\mathbf{X}_{\text{val}}$} and test set {$\mathbf{X}_{\text{test}}$}, batch size $\mathcal{B}$,

**Outputs:** Label for each sample

**Step 1. Model training**:

1. Initialize the model parameters.
2. **For** each training iteration from 1 to *T*:
    a)   Randomly select a batch of samples **x** from {$\mathbf{X}_{\text{train}}$} with size $\mathcal{B}$
    b)   If the model is GWVAE, sample the latent representation z by encoding the samples x using the SGC-encoder:
    $\mathbf{z}_\varphi^{(i)} \sim q_\varphi(\mathbf{z}| \mathbf{x}^{(i)})$ (i).
    c)   Update the SWGConv encoder and the feature decoder using Eq. (13) for GWAE or Eq. (14) for GWVAE.
3. **End for**

**Step 2. Fault detection:**

1. Estimate the threshold (Thr.) by evaluating the errors of $\mathbf{X}_{\text{val}}$ using Eq. 18.
2. Input the test set {$\mathbf{X}_{\text{test}}$} into the trained models to obtain the reconstruction errors.

3. Compare the reconstruction errors with the obtained Thr. to detect faults and estimate the AUC.

4. **Return the** Fault Labels and AUC.

## 4 Experiments

### 4.1 Datasets Collected on Fuel Control System

The fuel control system plays a crucial role in determining the control quality and performance of an aero-engine. Therefore, in order to validate the effectiveness of the proposed method in detecting faults in such mechanical systems, two datasets are collected from the fuel control system simulation test bench [38]. These datasets include the gear pump dataset and solenoid valve dataset. The test bench, depicted in Fig. 3(a), consists of two branches, where the first branch can be used to simulate the pump faults and the second branch can be used to simulate the valve faults. We conducted separate experiments on each branch to investigate the system's dynamics when a specific component fails. Table 1 provides information about the two datasets and further detailed descriptions of each dataset are provided in the following section.

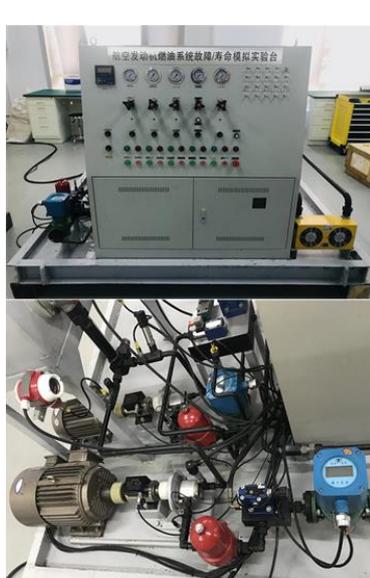

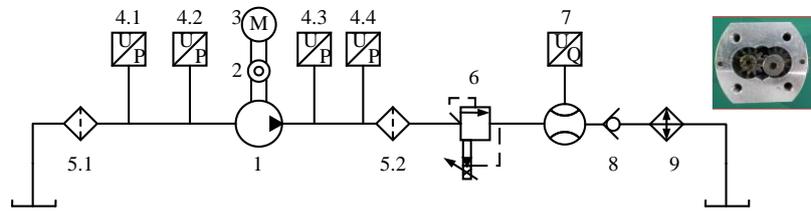

1. Gear pump; 2. Torque Tachometer; 3. Motor; 4.Pressure Sensor; 5.Oil filter; 6. Relief Valve; 7. Flow Meter; 8: Check Valve; 9: Cooler

(b) The first branch that fault component is the gear pump.

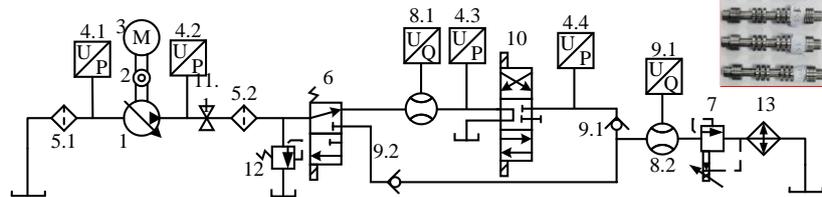

1. Plunger pump; 2. Torque Tachometer; 3. Motor; 4.Pressure Sensor; 5.Oil filter; 6.Solenoid reversing valve; 7.Proportional relief valve; 8.Flowmeter; 9.Check valve; 10.Solenoid valve; 11.Ball valve; 12.Safety valve; 13.Cooler.

(a) Fuel control system simulation test bench       (c) The second branch that the fault component is the solenoid valve

Fig. 3. The fuel control system test bench.

#### 4.1.1 Gear Pump Dataset

The schematic diagram for the gear pump fault simulation is shown in Fig. 3(b). In this experiment, we investigated the influence of the gear faults on the gear pump. As a result, eight types of gear faults were artificially created in the gear pump, including two types of tooth surface wear, three types of a broken tooth, and three types of floating sleeve end cover wear. During the experiment, one acceleration sensor mounted on the gear pump was used to collect the system condition. The data collection was performed at a sampling frequency of 20480 Hz, while the motor speed and the system pressure were set to 1500 rpm and 15 MPa, respectively. In this experiment, we consider the data collected from the gear pump in the aforementioned eight cases representing the abnormal state of the system. Furthermore, data is collected when the gear pump is functioning properly to represent the normal state of the system.

### 4.1.2 Solenoid Valve Dataset

The schematic diagram for solenoid valve fault simulation is depicted in Fig. 3(c), which exhibits greater complexity compared to the first branch (gear pump fault). In this experiment, we focus on investigating the impact of faulty or degraded solenoid valve on system's condition by intentionally introducing four different degrees of strain failure and four different degrees of wear failure of the solenoid valve spool. One acceleration sensor mounted on the solenoid is used to monitor the system status. During the experiment, the motor speed is set to 2000 rpm, and the system pressure is 8 MPa with a sampling frequency of 20480 Hz. Apart from the data collected in the aforementioned eight faulty conditions, we also collect data from the healthy system state.

Table 1. Description of two condition monitoring datasets.

| Dataset | Gear pump dataset | Solenoid valve dataset |
| --- | --- | --- |
| Fault component | gear | valve spool |
| Abnormal states | 8 | 8 |
| Training graphs/nodes | 40/400 | 40/400 |
| Validation graphs/nodes | 40/400 | 40/400 |
| Testing graphs/nodes | 180/1800 | 180/1800 |

## 4.2 Data Preprocessing and Implementations

### 4.2.1 Data Preprocessing and Graph Data Construction

The collected sensor measurements are first normalized using the max-min normalization technique,

which is defined as follows:

$$\bar{\mathbf{x}} = \frac{\mathbf{x} - \min(\mathbf{x})}{\max(\mathbf{x}) - \min(\mathbf{x})} \qquad (18)$$

where $\bar{\mathbf{x}}$ denotes the normalized sensor measurement, and $\min(\cdot)$ and $\max(\cdot)$ denote the minimum and maximum value of the training dataset, respectively.

After obtaining the sensor measurements, we proceed to split the sensor measurements into a set of subsamples that are required for the PathGraph construction using a sliding window approach. The resulting set of subsamples, denoted as $\Omega$, can be represented as follows:

$$\Omega = \{x_1, x_2, \ldots, x_N\}, N = \lfloor L/d \rfloor \qquad (19)$$

where $x_N$ denotes $N$-th sample, $L$ is the signal length, and $d$ is the length of sliding window. The notation $\lfloor \cdot \rfloor$ denotes rounding down.

With the subsample set $\Omega$, we can construct the PathGraph [21] for $n$ nodes to represent the temporal relationship between each node. In PathGraph, each node is sequentially connected and the edge weight is determined by:

$$A_{ij} = \exp\left(-\frac{\|(x_i - x_j)\|}{2B}\right), x \in \Omega_n \qquad (20)$$

where $B$ is the bandwidth of the Gaussian kernel. Here, we take the average distance between the neighboring nodes as $B$.

In this experiment, both the gear pump dataset and the solenoid valve dataset have a window length $d$ of 1024. The PathGraph is constructed for every 10 subsamples. This results in the generation of 100 graphs representing the normal state and 20 graphs for each abnormal state. Taking the signal in the normal state of the gear pump as an example, the constructed PathGraph is shown in Fig. 4. During the model training, 40% of the normal samples are randomly selected for both the training set and the validation set, where the validation set is also utilized to obtain the threshold. The remaining 20% of normal graphs, along with the graphs from abnormal states, are used as the test set. Therefore, each dataset consists of 40 graphs for model training and model validation, and 180 graphs for model testing. It is important to note that since we focus on a graph node-level learning task, there are in fact 400=40×10 training and validation samples, and 1800 test samples when performing FD, as demonstrated in Table 1.

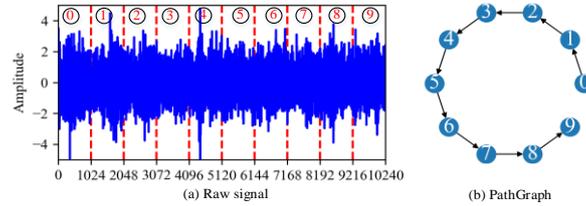

Fig. 4. An example of a PathGraph.

### 4.2.2 Comparison Models and Experimental Setup

In order to estimate the effectiveness of the proposed method, eight FD methods are used as comparison methods, including DeepSVDD, GraphSVDD, CAE, CVAE, GAE, GVAE and GUnet. The detailed structure of each method is listed in Table 2. Details of each applied method are provided in the following.

· **DeepSVDD.** DeepSVDD, introduced in [20] utilizes a DL model to learn the decision boundary between the normal and the abnormal data for FD. In this paper, we use a two-layer CNN as the feature extractor to learn the decision boundaries.

· **CAE.** CAE, as introduced in [18] utilizes the reconstruction error as the anomaly score and performs FD. Here, the number of linear layers for both the encoder and decoder is set to two.

· **CVAE.** CVAE, unlike CAE, is a generative model that reconstructs features from the reparametrized latent variables. Here, we follow the settings in [19] to perform fault detection.

· **GAE and GVAE.** GAE and GVAE are two unsupervised methods for modelling graph data, and their approach to FD is similar to AE and VAE [30]. Here, we align their structure with GWAE and GWVAE.

· **GraphSVDD.** GraphSVDD shares the same structure as DeepSVDD, but its CNN feature extractor is replaced with GVAE.

- **GUnet.** Graph U-net (GUnet) reconstructs the graph data using an encoder-decoder approach and maintains low-level features with a skip connection. In this research, we follow the settings in [39] and set the graph convolutional layers to 3 for both the encoder and decoder.

For model training, each model is trained for 100 epochs with the training dataset and the hyperparameters are optimized based on the validation dataset. Adam optimizer is used with an initial learning rate of 0.001. Weight decay is applied every 50 epochs. All the hyperparameters are keep the same for the proposed method, while the decomposition scale of SGWConv is set to 2. All the methods are written in Python 3.8 with Pytorch 1.11 machine learning framework and CUDA 1.11.3 version. The experiments are conducted on a computer with an Intel i7-11800H CPU and a GTX3080 GPU.

In these experiments, AUC is utilized as the primary performance evaluation metric, which is commonly used to evaluate the performance of anomaly detection approaches. Additionally, we calculate the fault detection accuracy (Acc) and F1score based on the estimated threshold as outlined in Section III. To reduce the randomness of the results, we repeat the experiments 5 times and report the average value of these experiments as the final results.

Table 2. The model structure of each method.

| Model | Structure |
|---|---|
| DeepSVDD [20] | **Encoder:** Conv1d(1,8)+Maxpool1d(2,2)+Conv1d(8,4)+Maxpool1d(2,2)+FC(28), |
| | **Decoder:** ConvT1d(2,4)+UPool()+ConvT1dT(4,8)+UPool ()+ConvT1dT(8,1) |
| GraphSVDD | **Encoder:** GConv(1024,1024)+ReLU()+GGonv(1024,512)+Gconv(1024,512), |
| | **Decoder:** FC(512, 1024)+ReLU()+FC(1024,1024) |
| CAE [18, 40] | **Encoder:** Conv1d(1,16)+ReLU()+Conv1d(16,32)+ReLU(), |
| | **Decoder:** ConvT1d(32,16)+ReLU()+ConvT1d(16,1) |
| CVAE [19] | **Encoder:** Conv1d(1,16)+ReLU()+Conv1d(16,32)+ReLU()+Conv1d(32,64)+ReLU()+Conv1d(64,128)+ReLU()+FC1(128*4,128)+FC21(128, 20)+FC22(128,20), |
| | **Decoder:** FC1(128*4,64)+ConvT1d(64,32)+ReLU()+ConvT1d(32,16)+ReLU()+ConvT1d(16,1) |
| GAE [30] | **Encoder:** GConv(1024, 1024)+ReLU()+GConv(1024, 512), |
| | **Decoder:** FC(512,1024)+ReLU()+FC(1024,1024) |
| GVAE [30] | **Encoder:** GConv(1024, 1024)+ReLU()+GConv (1024, 512)+GConv(1024,512), |
| | **Decoder:** FC (512,1024)+ReLU()+FC(1024,1024) |
| GUnet [39] | **Endoder:** GConv(1024,512)+ReLU()+TopKpool()+GConv(512,256)+ReLU(), |

| | | |
|---|---|---|
| GWAE | | **Decoder:** gUPool(512)+GConv(512, 1024) |
| | | **Encoder:** |
| | | SGWConv (1024,1024)+ReLU()+SGWConv(1024, 512), |
| | | **Decoder:** FC(512,1024)+ReLU()+FC(1024,1024) |
| GWVAE | | **Encoder:** |
| | | SGWConv(1024,1024)+ReLU()+SGWConv(1024,512)+ SGWConv (1024,512), |
| | | **Decoder:** FC (512,1024)+ReLU()+FC(1024,1024) |

## 4.3 Experimental Results

### 4.3.1 Analysis of results

The results are presented in Table 3, and the corresponding AUC curves can be seen in Fig. 5. From these results, it is evident that DeepSVDD and GraphSVDD can achieve excellent detection performance by learning decision boundaries. In contrast, the reconstruction-based FD methods show different performances, with GAE and GVAE outperforming CAE and CVAE. This suggests that incorporating structural information is advantageous for improving model performance. Furthermore, the proposed GWAE and GWVAE outperform all other fault detection models on both datasets, highlighting the superiority of the proposed SGWConv. Moreover, when compared to the best performing comparison method (GVAE and DeepSVDD), GWAE and GWVAE exhibit a significant improvement in AUC over GVAE by 3.87% and 3.39%, on gear pump dataset, and improve AUC over DeepSVDD by 4.12% and 4.27% on solenoid valve dataset. These promising results with respect to AUC values indicate the potential of the proposed methods for fault detection and also demonstrate consistently stable performance.

Table 3. AUC (%) for Gear Pump Dataset and Solenoid Valve Dataset.

| Dataset | DeepSVDD | GraphSVDD | CAE | CVAE | GAE | GVAE | GUnet | **GWAE** | **GWVAE** |
|---|---|---|---|---|---|---|---|---|---|
| Gear pump | 94.86±1.13 | 93.54±0.11 | 85.92±0.98 | 88.69±0.24 | 95.60±0.28 | 95.91±0.63 | 89.37±0.81 | **99.78**±0.15 | 99.30±0.06 |
| Solenoid valve | 93.44±1.12 | 87.27±1.46 | 91.79±0.75 | 92.21±0.25 | 90.00±0.51 | 93.33±0.03 | 93.02±0.04 | 97.56±0.01 | **97.71**±0.06 |

In addition, we also visualized the distributions of the reconstruction errors of the proposed GWAE and GWVAE in Fig. 6, which are estimated by KDE. It can be observed that the distribution of the training and the test sets of normal samples is largely overlapping, but they can be effectively distinguished from abnormal samples. This demonstrates that GWAE and GWVAE are capable of learning discriminative features to effectively differentiate between normal and abnormal samples.

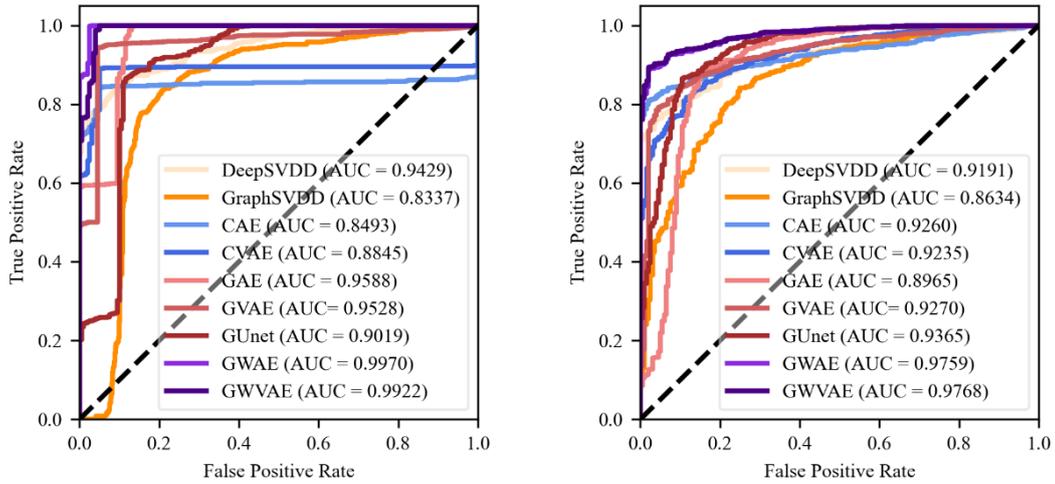

T

Fig. 5. The AUC curves of all the applied methods of the two datasets. (a) gear pump dataset, and (b) solenoid valve dataset.

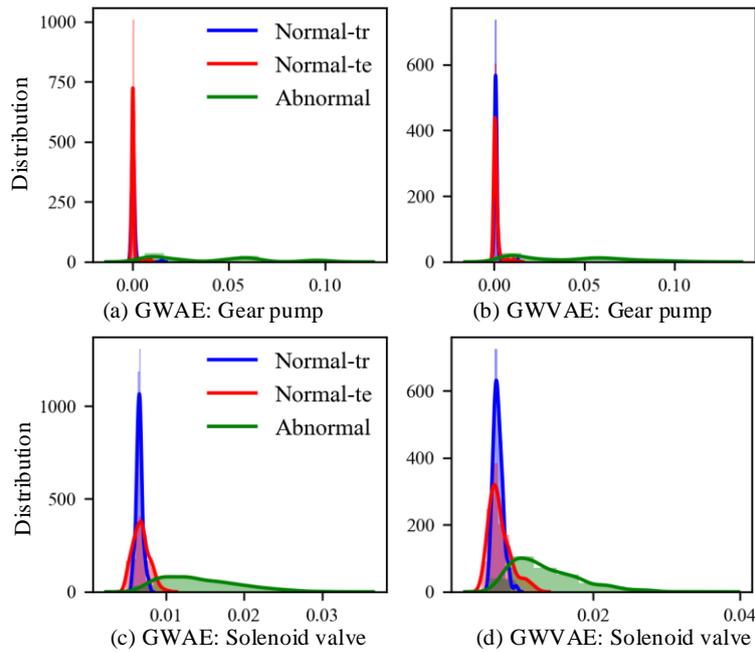

.

Fig. 6. The statistical distribution of the reconstruction errors of GWAE and GWVAE on gear pump dataset, and solenoid valve dataset. (Normal-tr: normal samples for model training; Normal-te: normal samples for model testing, Abnormal: abnormal samples for model testing.)

### 4.3.2 Fault Detection with the Obtained Threshold

Using the obtained reconstruction errors of each model on the validation dataset, we can estimate the decision thresholds (Thr.) using Eq. 18. Additionally, based on the obtained threshold, we can calculate the FD accuracy and the F1score of each model on the test set, which are presented in Table 4.

Table 4. FD Acc (%), F1 Score (%) and Threshold (Thr.) for Gear Pump and Solenoid Valve Dataset.

| Model | Gear pump | | Thr. | Solenoid valve | | Thr. |
|---|---|---|---|---|---|---|
| | Acc | F1score | | Acc | F1score | |
| DeepSVDD | 93.18±1.39 | 95.64±0.93 | 0.1743 | 90.74±1.01 | 94.98±0.56 | 0.0752 |
| GraphSVDD | 92.06±1.65 | 93.49±0.62 | 1.5696 | 88.43±0.84 | 93.72±0.16 | 13.695 |
| CAE | 82.25±0.85 | 89.75±1.17 | 0.0016 | 88.46±1.93 | 93.26±0.92 | 0.0022 |
| CVAE | 83.70±0.74 | 87.60±1.27 | 0.0261 | 91.67±0.39 | 95.00±0.10 | 0.0171 |
| GAE | 94.42±1.45 | 95.12±2.02 | 0.0016 | 91.78±0.49 | 95.38±0.43 | 0.0066 |
| GVAE | 90.23±1.49 | 94.21±0.66 | 0.6413 | 90.95±0.16 | 94.97±0.07 | 0.6544 |
| GUnet | 90.13±1.23 | 92.28±0.87 | 0.0007 | 91.18±0.19 | 95.28±0.11 | 0.0062 |
| **GWAE** | **97.74±1.05** | **98.38±0.27** | 0.0002 | 92.20±0.09 | 95.79±0.05 | 0.0062 |
| **GWVAE** | 96.07±0.29 | 97.84±0.13 | 0.0008 | **92.70±0.04** | **95.99±0.09** | 0.0066 |

As observed from Table 4, the proposed GWAE and GWVAE consistently achieve the highest detection accuracy and F1score on both gear pump and solenoid valve dataset. Specifically, when comparing the accuracy of GWAE and GWVAE with GAE using the obtained threshold, they exhibit improvements of 3.32% and 1.65% respectively on the gear pump dataset. Similarly, on the solenoid valve dataset, GWAE and GWVAE exhibit improvements of 0.42% and 0.92% respectively compared with GAE. The F1scores of GWAE and GWVAE on gear pump dataset surpass DeepSVDD by 2.74% and 2.2% respectively. Similarly, on the solenoid valve dataset, they exhibit improvements of 0.81% and 1.01% respectively compared to DeepSVDD. This high fault detection accuracy and F1scores further validate the effectiveness of the proposed methods.

### 4.3.3 The Influence of Decomposition Scales

To investigate the impact of the decomposition scale of SGWConv on model performance, we conducted experiments by incrementally increasing the scale parameter from 2 to 10. The results are depicted in Fig. 7.

From these results, we can observe that the Acc and F1score of each model remain relatively stable as the decomposition scale increases, with a deviation bounded within ±1.5%. However, it is noteworthy that the AUC values of GWVAE exhibit a rapid decline when the decomposition scale is exceeds 6. This suggests that excessively large decomposition scales are unfavorable for generative models to perform effective variational inference. Therefore, it is advisable to select a smaller decomposition scale in practical applications.

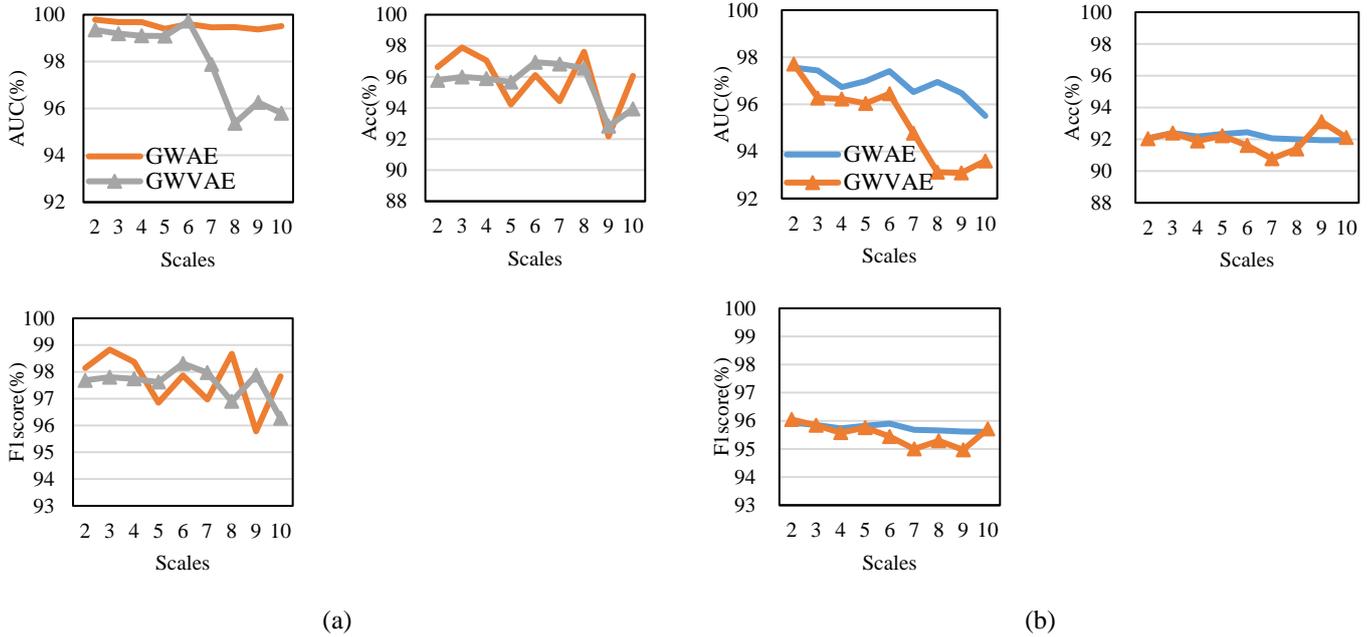

Fig. 7. The influence of the decomposition scale. (a) gear pump dataset, and (b)solenoid valve dataset.

## 5 Fault detection under noisy conditions

### 5.1 Data description

To assess the effectiveness of the proposed methods in detecting faults of malfunctioning industrial machines under noisy conditions, we conducted experiments on the valve dataset, which is a subset of the MIMII dataset [41]. This dataset consists of sound signals recorded from industrial valves, including signals from the normal and abnormal states under three different signal-to-noise ratio (SNR) setups (6 dB, 0 dB and -6dB). Only healthy data is available for training. Faulty samples are only contained in the testing dataset. During the experiments, eight microphones are utilized to collect the sound signals at the sampling frequency of 16 kHz and only data from the first sensor are utilized. Specifically, we focused on two valve datasets identified as #ID00, #ID02 to evaluate the performance of the models. Detailed information about these datasets is provided in Table 5.

### 5.2 Experimental results

In our experiments, we focused on the signals from the first channel, following a similar approach as previous experiments. Each sample was converted into the PathGraph representation, with a node feature dimension of 1024. Additionally, the remaining experimental setup was consistent with the previous experiments. The corresponding results are presented in Table 6. The ROC curves at -6dB are illustrated in Fig. 8.

Table 5. Details of the valve dataset.

| Data ID | Normal samples | Abnormal samples | Operation conditions and example of anomalous conditions |
|---|---|---|---|
| 00 | 991 | 119 | Open / close repeat with different timing. |
| 02 | 708 | 120 | More than two kinds of contamination |

It can be observed from the results that the proposed methods outperformed all the comparison methods. Specifically, on dataset #ID00, GWAE and GWVAE achieved an average AUC improvement of 1.49% and 1.62%, respectively, compared to the best performing comparison method GAE. On dataset #ID02, GWAE and GWVAE exhibited improvements of 8.23% and 7.65% respectively compared to GVAE. Besides, at each noise level, both GWAE and GWVAE can obtain the smallest standard deviation, which demonstrates consistently stable and robust performance.

To analyze the robustness of each method at different noise levels, we also calculate the average AUC of each method on two datasets under the same SNR. The results are shown in Fig. 9. It can be observed that when SNR changes from 6 dB to 0 dB, the AUC of GWAE and GWVAE only exhibit a change of 0.65% and 0.72% respectively. While DeepSVDD, GraphSVDD, CAE, CVAE, GAE, GVAE and GUnet relatively change 4.27% 0.7%, 5.2%, 10.02%, 3.24%, 2.05%, and 1.95%, respectively. Similarly, when the noise increases from 0 dB to -6 dB, the performance of GWAE and GWAVE changes by only 0.36% and 0.64%, respectively. These results indicate that the proposed methods can effectively capture the noise-independent features from raw signals, thereby enhancing robustness in the presence of noise compared to other methods. This attribute is crucial in practical applications as real-world applications often involve various types and levels of noise.

Table 6. AUC (%) of each method on the valve dataset.

| Dataset | SNR | DeepSVDD | GraphSVDD | CAE | CVAE | GAE | GVAE | GUnet | **GWAE** | **GWVAE** |
|---|---|---|---|---|---|---|---|---|---|---|
| #00 | 6dB | 93.54±0.60 | 85.42±1.57 | 94.37±0.67 | 94.27±0.53 | 94.56±0.39 | 93.95±0.65 | 82.08±0.87 | **97.15±0.37** | 96.89±0.37 |
|  | 0dB | 93.17±0.53 | 83.75±0.47 | 96.08±0.53 | 96.18±0.26 | 95.93±0.20 | 94.94±0.31 | 88.86±1.61 | 97.06±0.42 | **98.17±0.32** |
|  | -6dB | 93.03±0.36 | 83.77±1.33 | 97.67±1.32 | 96.68±0.57 | 97.81±1.13 | 97.62±0.53 | 94.12±0.34 | **98.84±0.28** | 98.81±0.30 |
|  | Avg. | 93.25±0.49 | 84.31±1.12 | 96.04±0.84 | 95.71±0.45 | 96.10±0.57 | 95.50±0.49 | 88.35±0.93 | 97.68±0.35 | **97.96±0.33** |
| #02 | 6dB | 73.67±0.39 | 84.95±0.35 | 81.15±0.91 | 83.47±0.59 | 89.89±0.35 | 95.31±0.55 | 54.86±1.88 | **95.23±0.21** | 95.17±0.37 |
|  | 0dB | 65.50±2.16 | 85.23±0.71 | 89.84±0.22 | 61.51±0.60 | 82.05±1.80 | 90.22±0.56 | 51.98±0.62 | **96.62±0.35** | 95.33±0.14 |
|  | -6dB | 59.07±3.44 | 85.04±0.59 | 83.54±1.25 | 65.27±1.03 | 79.59±1.54 | 76.17±0.76 | 49.42±0.91 | **94.40±0.18** | 94.13±0.39 |
|  | Avg. | 66.08±1.99 | 85.07±0.55 | 84.84±0.79 | 70.08±0.73 | 83.84±1.151 | 87.23±0.62 | 52.09±1.13 | **95.42±0.24** | 94.88±0.30 |

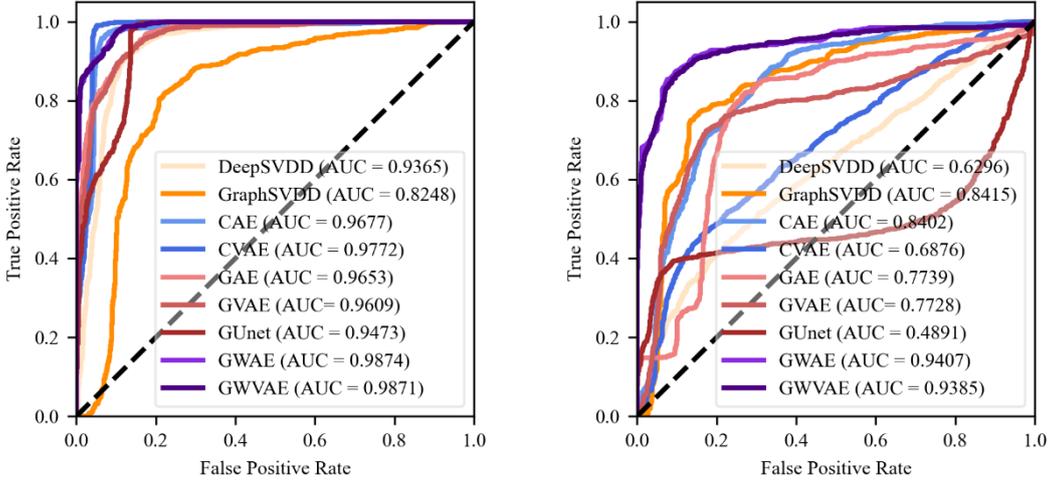

Fig. 8. The AUC curve of valve at -6dB. (a) #00 dataset, (b) #02 dataset.

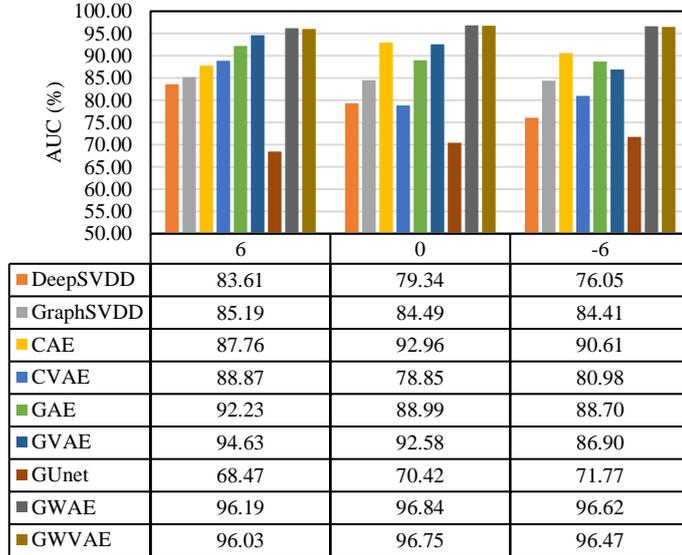

Fig. 9. The average AUC of each method at different SNR.

To further explore the ability of the proposed methods to distinguish between normal and faulty samples, we visualized the distributions of the reconstruction errors of the proposed GWAE and GWVAE in Fig. 10 and Fig. 11. As can be seen from these figures, the learned distribution centers of normal training samples and the normal test samples are essentially the same, whereas the distribution center of the normal test samples deviates from the abnormal samples. This illustrates why the proposed methods can obtain a good AUC under different noise levels. However, we must acknowledge that there are still overlaps between the distribution of normal test samples and abnormal samples, indicating the potential for further improvement in our future work.

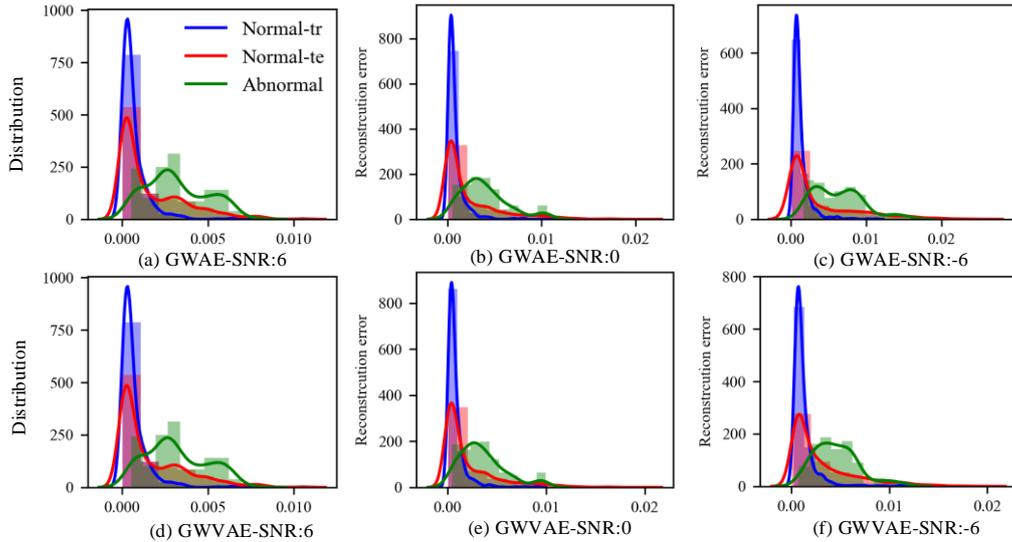

Fig. 10. The statistical distribution of the reconstruction errors of GWAE and GWVAE (Dataset #00 test).

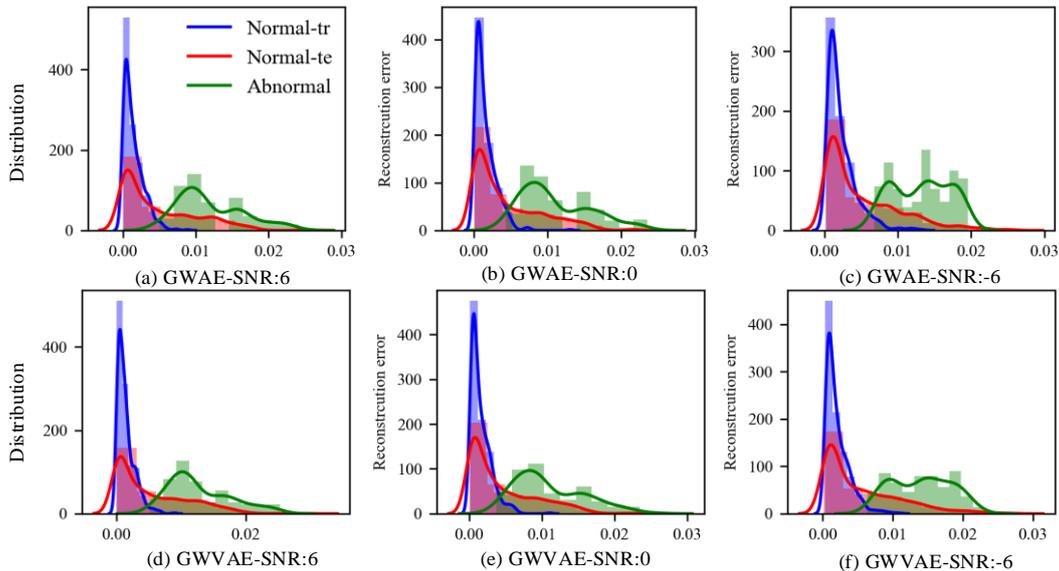

Fig. 11. The statistical distribution of the reconstruction errors of GWAE and GWVAE (Dataset #02 test).

## 5.3 Comparison with existing literature results

Moreover, we also compare the results of the proposed methods to the results found in previous research studies, including the baseline (AE) results reported in [41], DAE [42], ConvAE [42], LAE [43], dictionary learning (DicL) [44], L3music [45], DenseNet with Image input (DenseImag) [45], and KNN using CNN features as input (CNNKNN) [46]. These results are listed in Table 7.

It can be observed from Table 7 that the proposed methods outperform the comparison methods and achieve the best performance. When compared to the baseline, GWAE and GWVAE show a significant overall improvement of 34.85% and 34.58%, respectively. Furthermore, when compared to the best-performing method (CNNKNN and DicL), GWAE and GWVAE still achieve improvements of 1.01% and 1.14% on dataset #ID00, and improvements of 7.52% and 6.98% respectively on dataset #ID02. These results further demonstrate the superiority of the proposed method.

Table 7. AUC (%) reported in the exiting literatures.

| Dataset | Baseline[41] | DAE[42] | ConvAE[42] | LAE[43] | DicL[44] | L3music[45] | DenseImag[45] | CNNKNN[46] | **GWAE** | **GWVAE** |
|---|---|---|---|---|---|---|---|---|---|---|
| #00 | 62 | 74.61 | 78.69 | 70.99% | 54.00 | 74.53 | 84.16 | 96.58 | 97.59 | **97.72** |
| #02 | 66 | 76.68 | 85.02 | 66.75% | 87.90 | 78.93 | 69.10 | 52.78 | **95.42** | 94.88 |

# 6 Conclusion

In this paper, we propose two novel graph neural networks, GWAE and GWVAE, based on SGWConv, specifically designed for mechanical system fault detection. These models consist of an SGWConv encoder and a feature decoder, with GWVAE being the variational form of GWAE. Moreover, the SGWConv is defined by the spectral graph wavelet transform allowing for multiscale feature extraction. For fault detection in mechanical systems, the proposed approach involves transforming the collected system signals into PathGraph representation to capture temporal relationships at different scales in the high-frequency condition monitoring data. The graph neural networks then perform graph representation learning and feature reconstruction. By analyzing the reconstruction error of normal and abnormal data, we achieve fault detection for mechanical systems. We conduct experiments on two datasets collected from fuel control systems and an acoustic monitoring dataset of valves. The results demonstrate superior performance of our proposed methods compared to the comparison methods, validating their effectiveness.

In future work, we will explore the application of our proposed methodology to fault detection of low-frequency signals and aim to achieve transferable anomaly detection through adversarial training.

## Acknowledgments

This work was supported in part by Natural Science Foundation of China (No. 52175116) and Major Research Program of Natural Science Foundation of China (No. 92060302). The work of Tianfu Li was also supported by the China Scholarship Council (CSC) for one year's study at the École Polytechnique Fédérale de Lausanne (EPFL).